\documentclass[12pt,preprint]{aastex}
\shorttitle{The luminosity function of the LMC cluster NGC~1866}
\shortauthors{Brocato E. et al.}

\newcommand{\lsim}{\ \raise
-2.truept\hbox{\rlap{\hbox{$\sim$}}\raise5.truept\hbox{$<$}\ }}
\newcommand{\gsim}{\ \raise
-2.truept\hbox{\rlap{\hbox{$\sim$}}\raise5.truept\hbox{$>$}\ }}

\begin{document}

\title{The luminosity function of the Large Magellanic Cloud globular cluster NGC~1866}
\author{Brocato, E.\altaffilmark{1}, Castellani, V.\altaffilmark{2,3},
Di Carlo, E.\altaffilmark{1}, Raimondo, G.\altaffilmark{1,4}, and Walker A. R.\altaffilmark{5}}

\altaffiltext{1}{INAF - Osservatorio Astronomico di Teramo, via M. Maggini, I-64100 Teramo, Italy; brocato, dicarlo, raimondo@te.astro.it}
\altaffiltext{2}{INFN - Sezione di Ferrara, via Paradiso, 12, I-44100 Ferrara, Italy}
\altaffiltext{3}{INAF - Osservatorio Astronomico di Roma, Via di Frascati, 33, I-00040 Monte Porzio Catone, Italy; vittorio@mporzio.astro.it}
\altaffiltext{4}{Dipartimento di Fisica, Universit\`a La Sapienza,  P.le A.
Moro 2, I-00185 Roma, Italy}
\altaffiltext{5}{Cerro Tololo Inter-American Observatory, National Optical Astronomy Observatory, Casilla 603, La Serena, Chile;
awalker@noao.edu. NOAO is operated by AURA
Inc., under cooperative agreement with the National Science Foundation.}

\begin{abstract}

We present {\it Hubble Space Telescope} {\it V,I} photometry of
the central region of the LMC cluster NGC~1866, reaching
magnitudes as faint as $V=27$ mag.  We find evidence that the
cluster luminosity function shows a strong dependence on the
distance from the cluster center, with a clear deficiency of low
luminosity stars in the inner region.  We discuss a {\it global}
cluster luminosity function as obtained from stars in all the
investigated region, which appears in impressive agreement with
the prediction from a Salpeter mass distribution. We also revisit
the use of NGC~1866 as a probe for determining the efficiency of
core overshooting, and conclude that a definitive answer to this
question is not possible from this cluster.

\end{abstract}


\section{Introduction}

NGC~1866 is a Large Magellanic Cloud (LMC) stellar cluster which
is an extragalactic counterpart of young galactic clusters like the
Pleiades, but with a much richer stellar population.  It is probably
one of the most massive objects formed in the LMC during the last $3
\, Gyr$, with its total mass estimated as $M_{tot} = (1.35 \pm 0.25)
\times 10^5\, M_{\sun}$ by Fischer et al. (1992). Because of that
richness, it presents a statistically significant sample of He-burning
giants (more than 100 compared to zero in the Pleiades), allowing the
testing of predictions by stellar evolution models concerning late
evolutionary phases in intermediate mass stars, i.e., in stars with
masses of the order of $\sim 4-5 \, M_\odot$ (Arp \& Thackeray 1967;
Becker \& Mathews 1983: BM83; Chiosi et al. 1989: C89; Brocato et al. 1989: B89;
Testa et al. 1999: T99; Barmina, Girardi \& Chiosi 2002: BGC02).

The cluster dynamical mass has recently been investigated by van den
Bergh (1999), who by a comparison of the cluster integrated photometry
with the velocity dispersion measured by Fischer et al. (1992),
 found a very high mass-to-luminosity ratio ($M$/L$_V = 0.42$ in solar
units), suggesting the possible occurrence of an unusual mass spectrum, much
steeper than Salpeter's distribution (Salpeter 1955),
and concluding that {\it "clearly it
would be of great interest to obtain deep {\it
Hubble Space Telescope (HST)} photometry of this
cluster to confirm this conclusion"}.

At about the same time we were obtaining {\it V} and {\it I} {\it HST}
images of the cluster, aiming to extend previous investigations to
fainter magnitudes as well as to reveal for the first time the stellar
population right to the cluster center. In a previous paper (Walker et
al. 2001: Paper I) we presented the first results of this {\it HST}
photometry, discussing the Color-Magnitude (CM) diagram location of
the cluster Main Sequence (MS) in connection with the problem of the
LMC distance modulus. In the present paper we will use data corrected
for completeness and field contamination to derive a new stellar
luminosity function that extends to fainter than $V=25$ mag. In
section 2 we present the data reductions and the derived luminosity
function is studied in section 3. The constraints on the Initial Mass
Function (IMF) are discussed in section 4.  Section 5 is devoted to a
re-examination of the problem of overshooting in modelling the
convective cores of stars.  Final remarks close the paper.

\section{The observations}

The WFPC2 images of NGC~1866 were collected as target of the {\it
Hubble Space Telescope} program GO-8151. Two telescope pointings were
obtained, the first with the cluster core centered on the PC
(thereinafter: PC), the second with NGC~1866 at the center of the WF3
camera (WF3). This choice allows us  to perform two independent data
reductions, and to make internal checks on the reproducibility and reliability
of the results. The exposure times for the two selected filters {\it
F555W} and {\it F814W} are given in Table 1, together with
other observation details.  The three sets of exposure times allow
us to cover the high dynamic range of brightness of the cluster stars
with good overlap between sets.

The WFPC2 mosaics of the two fields are presented in Fig.~1 as
observed in the medium exposures images with filter {\it F814W},
whereas Fig.~2 compares the present PC-centered fields with fields
covered in previous ground based observations.

The removal of cosmic rays and the photometry has been performed using
the package HSTphot as developed by Dolphin (2000a) and according to
the recipe outlined by the author.  The most recent version of the
HSTphot package has been used, which allows the simultaneous
photometry of a set of images obtained by dithered multiple exposures,
as is our case.  Each frame has been preprocessed according to the
standard {\it HST} pipeline making use of the latest available
calibrations which are expected to provide the most accurate and stable
results.

Following Dolphin (2000a) we used the subroutine {\it mask} to
take advantage of the accompanying data quality images by removing
bad columns and pixels, charge traps and saturated pixels.  The
same procedure is also able to properly solve the problem of the
vignetted regions at the border of the chips.  The {\it crmask}
subroutine uses the available frames to clean the cosmic rays.

One of the advantages of HSTphot is that it allows use of PSFs which are
computed directly to reproduce the shape details of star images as
obtained in the different regions of the WFPC2.  For this reason we
adopt the PSF fitting option on the HSTphot routine, rather than use
aperture photometry.

A particular difficulty in attempting to do photometry on these frames
is the detection of fictitious objects,  mainly located around the bright, very
saturated stars on the medium and long exposures. These objects
resemble the faint stars we are trying to measure, therefore we
developed a procedure to get rid of the former, as follows.  We selected
all the stars located along {\it suspected} curves, as given by
$i)$ circular distribution of objects around ($d\leq 30$ pixels)
bright-saturated stars and, $ii)$ alignments (within 2 pixels)
along lines centered on bright-saturated stars with an angle of 45
degrees and at distances less than 49 pixels.
Our algorithm rejected $7 \%$ of objects on the WF3
frames and $6 \%$ on the PC-centered frames.

Additionally, the HSTphot photometry routine returns various data
quality parameters which can aid in the removal of spurious
objects, we found the most useful of these to be the object
classification parameter which, in addition to the objects
rejected by our algorithm as described above, found $13 \%$ of
objects for the WF3-centered and $8 \%$ for the PC-centered frames
to be spurious. The sharpness parameter would mainly find objects
fainter than $V \sim 25$, presumably faint galaxies and stars with
low S/N,
while the $\chi $ and roundness parameters do not significatively
improve the photometric results.  We find that using our method
combined with the HSTphot object classification parameter is a
very effective method to identify fictitious objects.

In conclusion, by analyzing the list of objects provided by the
output files of HSTphot we find that the percentage of identified
spurious objects is $20 \%$ and $14 \%$ for the WF3-centered and
PC-centered data respectively.  This difference is not surprising
since the intensity of the saturation effects, the major effect
responsible for generating spurious objects, depends on the
exposure times which are different between the two sets of data.

The results of all these selections are shown on Fig.~3 where the
CM diagrams of the rough photometry are compared to the CM
diagrams as obtained after our selection.  The CM diagrams show
all the evolved stars located in the observed area.  Within the
box defined as $V=17.4-15.2$ and $0.3 \leq (V-I)\leq (23.7-V)/5.$
we find a number of red giant
(RG\footnote[1]{We are aware that the box also includes
 'blue' stars populating the loop region during the core
He-burning phase. In this paper we will use the terms
"RG stars" and "He-burning giants" interchangeably.}) stars
$N_{RG}=128$ and $N_{RG}=153$, respectively for the PC-centered
and WF3-centered dataset.
 We also note that the CM diagram clearly shows the presence
of numerous evolved field giants in the LMC itself.

Charge Transfer Efficiency (CTE) corrections and calibrations to the
standard $VI$ system were obtained directly by HSTphot routines, as
documented by Dolphin (2000b). Fig.~4 shows the uncertainties of
photometry as a function of the magnitude for the two
filters. Completeness has been evaluated by distributing artificial
stars of known positions and magnitudes, in selected circular regions
around the cluster center. The adopted procedure allows the
distribution of stars with similar colors and magnitudes as in the
real CM diagram. The resulting completeness functions are shown in
Fig.~5 for the two sets of data and for the two selected filters. As
expected, in both samples WF3 and PC the limiting magnitude for
completeness decreases when moving from the periphery toward the
cluster center.

To estimate the contamination by Galactic foreground stars, we
used the tabulation by Ratnatunga \& Bahcall (1985). We expect
negligible MS contamination. Only 1.5 Galactic foreground stars
are expected to be present in the region of the field red clump,
while only 0.8 in the cluster RG region.  Thus we conclude that
most contamination is due to LMC field stars.

The LMC field contamination has been evaluated following Stappers
et al. (1997). They observed the region around NGC~1866 in an area
of $668\,arcmin^2$ and derived the star counts of field stars in
the proximity of NGC~1866.  We normalized the field stars numbers
to the area covered by the 3 WF chips plus the PC chip for a total
of $5\,arcmin^2$, after the rejection of the pixels from 0 to 51.5
for each camera on average, as suggested by the {\it WFPC2}
handbook manual.  We note that for $V$$<$17 mag the field
contamination is very small, thus supporting the fact that
the He-burning giants ($N_{RG}$) quoted above refers to cluster members
(within $1\sigma$).

 The contamination of the cluster main sequence by the MS
stars in the LMC field has been evaluated by following the same
procedure. Again, we find that less than 1-2 field stars may
contaminate the turn-off region. This ensures us that the
conclusions about the age and evolution of the stars, obtained on
the basis of the CMDs presented in this work, are not affected by
LMC field stars.  Since the number of field stars significatively
increases towards fainter magnitudes, this contamination directly
affects the luminosity function of the lower part of the main
sequence. For this reason we used the rescaled data by Stappers et
al. (1997) to evaluate the number of LMC field stars that have to
be subtracted in each bin of the luminosity functions derived from
the original CMDs.

To evaluate the number of background faint galaxies that may affect our
luminosity function we adopt the data provided by Metcalfe et al. (2001).
According to their work on the Hubble Deep Fields, we
expect a number of faint galaxies smaller than the reported
uncertainties ($1 \sigma$) in each bin of the luminosity function of
NGC~1866 brighter than V $\simeq$ 24.5.  In particular, for V$\simeq$
24.5 (bin of 0.2 mag) the number of expected galaxies in our field is of
the order of $\sim$ 15 while the measured stars total 450 ($\pm
20$). For this reason in the following calculations we will disregard this
galaxy contamination.

In summary, all the LFs used in the following sections have been
corrected for spurious objects, completeness and LMC field star
contamination.  We have investigated possible contamination by
galactic field stars and faint galaxies, and in both cases found their
contributions to be negligible.

\section{The luminosity function}

We first discuss the luminosity function (LF) of the PC-centered
dataset which provides firm information,
particularly on the inner core region, unaffected by crowding at least
down to $ V \sim 23 $ mag. This limit is well below the
magnitudes reached in
the previous deepest ground-based observations
(T99), thus allowing a
substantial improvement in our knowledge of the cluster luminosity
function.
Fig.~6 shows the luminosity function of the cluster MS as
obtained in different annuli
around the cluster center. It discloses that the shape of the LFs
shows a strong dependence on the distance from the
cluster center, with a clear deficiency of low luminosity stars in the
inner region. Moreover, since data in the quoted figure give the
distribution of MS stars as normalized to the corresponding number of
He-burning giants, one finds also clear evidence for an increasing
abundance of giants when going toward the cluster center.

The problem of the origin of this observed mass segregation in NGC~1866
is beyond the scope of this paper. However, we note that strong evidence for
mass segregation has recently been found by de Grijs et al. (2002)
 in six LMC clusters of different ages,
possibly as a result of the clusters formation process.
In the case of NGC~1866, Fischer et al. (1992)
found a half-mass relaxation time
larger than $3\, Gyr$ for $M$$<$$0.75\, M_\odot$.  By following the same
approach one finds a half-mass relaxation time for the more massive
stars ($\simeq 4\, M_{\odot}$) of the order of $700\, Myr$, i.e., still
larger than the estimated cluster age (i.e. $\sim 140\, Myr$ see section~4),
and with a core relaxation time
of about $30\,Myr$.
Thus, the observed segregation should be largely
the result of an original segregation because not enough time has
elapsed for relevant dynamical mass segregation outside the core,
i.e. $r_c \gsim 14\, arcsec$.
It is worth mentioning that evidence of (dynamical)
mass segregation has been found in the old Galactic globular clusters
 (e.g. Albrow et al. 2002).

Given such an inhomogeneous spatial distribution, the problem
arises how to define a luminosity function for the whole cluster.
In the remainder of this paper we assume that the mass segregation
observed has only the effect of redistributing masses locally in
the cluster without affecting the overall cluster luminosity
function (see also de Grijs et al. 2002). Fig.~2 shows that the
PC-centered frames integrated with the region sampled by the
WF3-centered frames cover quite a large region of the cluster,
extending well beyond the half-mass radius ($ \sim 50\, arcsec$,
Fischer et al. 1992).  Thus in the following we will take the
luminosity function of such a sample as reasonably representative
of the whole cluster.  This LF (labelled as {\it global}) is shown in
Fig.~7 in comparison with the LF of the PC-centered and the
WF3-centered datasets alone. The total number of RG stars is then
$N_{RG}=170$.

Errors in the counts of stars in each bin of magnitude have been
computed accordingly to the following relation by Bolte (1989):
\begin{equation}
\sigma^{2}_{n} = [\frac{N_{obs}}{\lambda_{i}^2} + \frac{\sigma_{\lambda_{i}}^{2} \times N_{obs}^{2} }{\lambda_{i}^{4}}]
\end{equation}
where $\lambda_{i}$ are the completeness factors and $N_{obs}$ is the
star counts in the $i-th$ $V$-bin. This formula properly considers the
uncertainty due to the completeness as shown in Fig.~7 for each bin
value.  The error due to the statistical fluctuation of the RG stars
number ($\sqrt { N_{RG}}$) has also to be taken into account when
dealing with differential LF normalized to the RG stars.  This latter
error is that reported at the right bottom of the Fig.~7.

Comparison with previous work in the literature, as given
in Fig.~8, discloses that the present LF, when normalized to the
observed number of RG stars, tends to be less populated, with a
difference which increases, going toward less luminous stars.  This
appears to be directly related to the fact that the present
photometry, for the first time, measures stars to the core of the
cluster, allowing investigation of the LF in this internal region.

\section{Theoretical constrains on the IMF}

In order to obtain a preliminary indication on the cluster age, the
left panel in Fig.~9 compares the cluster CM diagram with the
predicted distribution of star for selected assumption about the age,
as evaluated assuming classical stellar models with inefficient
overshooting.  As in Paper I, stellar models were computed adopting
the evolutionary code developed for the stellar library under
construction at the University of Pisa (Degl'Innocenti et al., in
preparation).  The adopted physics input has been exhaustively
described in Cassisi et al (1998).
From the results of Paper I we
assume for cluster stars $Z=0.007$, $Y=0.25$, $(M-m)_V=18.55$.

From data in Fig.~9 one finds that a reasonable fit of the He
burning giants requires an age of about $140\, Myr$. To be
conservative, in the following we will adopt as a safe estimate of
the cluster age the interval $t= 100 - 180 \, Myr$. The same
comparison is given in right panel of the same figure, but for
models allowing a moderate overshooting (by $0.25\,H_P$).  As is
well known, one finds that increasing the amount of overshooting
increases the estimated cluster age, and a reasonable fit now
requires an age of the order of $200\, Myr$. We will assume
the interval $t=160 - 250\, Myr$ as a safe estimate of the actual
cluster age if overshooting is at work. It is important to note,
as found from previous similar investigations (T99, BGC02),
that the age from the MS termination appears in all
cases (i.e.  with or without overshooting) lower than the age
derived from the fitting of the luminosity of the RG stars. This
is generally taken as evidence for the occurrence of a substantial
number of binary stars which apparently increase the limiting
luminosity of the observed MS. Table~2 gives the predicted mass of
He-burning giants for the various cases, together with the mass of
MS stars at two selected luminosities (see later).

In order to discuss the predicted luminosity functions we will adopt
everywhere a logarithmic representation, which allows removal of the
effect of the sample abundance, which only produce a shift along the
Y-axis, keeping untouched the overall shape and slope of the
distribution (Castellani, Chieffi, \& Norci 1989).  Fig.~10 shows the
predicted luminosity distribution of a suitable sample of MS stars, as
evaluated with or without overshooting, a Salpeter IMF and for cluster
ages covering the quoted safe intervals derived from the luminosity of
He-burning giants.  One finds that -- in both cases (with or without
overshooting) and within the safe range of ages -- age can affect
the distribution only above $M_V \sim 0$ whereas for fainter
luminosities the distribution is determined only by the IMF. Note
that, in this respect, the differential LF (DLF) we are dealing with,
is superior to the often adopted ``cumulative'' LF (CLF), where the (age
dependent) upper portion of the DLF affects the cumulative
distribution even at the lower luminosities.

The predicted sensitivity to the IMF is shown in Fig.~11  where
we show the behavior of the DLF for a given age but for
selected assumptions about the IMF exponent for classical models. As
expected, one finds that assumptions about the IMF clearly modulate
the behavior of the LF fainter than $M_V \sim 0$.  For the given
distance modulus, this means that in the magnitude interval 18.55 to
23 mag we  directly explore the cluster IMF over an interval of
more than 4 magnitudes without the need of corrections for the
completeness of the sample. In such a way we derive information
on the IMF between $M_V=0$ and 4.5, approximately corresponding to the
mass range $M= 1$ to $3.5\, M_{\sun}$, reaching a mass of about
$0.7\, M_{\sun}$ at the limiting magnitude $V\sim 25$.

The best fit to theoretical predictions, as given in the same figure,
gives the clear evidence that in the explored range of luminosities
the cluster IMF closely follows a Salpeter law.  It follows that the
suggestion for a larger slope, as proposed by van den Bergh (1999), is
not supported by observational data, at least in the explored range of
stellar masses.  As quoted before, the present results rely on the
hypothesis that the mass segregation present in this cluster affects
the local mass distribution but not the whole cluster mass function.

Let us note that a similar result for the behaviour of the IMF is also
found by de Grijs et al. (2002) for a sample of six LMC clusters.  As
in our case, they found a radial dependence of the luminosity function
slope (then mass function), as the direct consequence of the mass
segregation, the overall cluster IMF slope was found to be close to a
Salpeter slope.

The fact that the luminosity function of NGC 1866 agrees with a
Salpeter mass distribution clearly affects the evaluation of the
mass-to-light ratio ($M$/L$_V$).  As a first point, one may recall
that previous computations of the dynamical mass-to-light ratio were
forced to adopt the mass function distribution as a nearly free
parameter due to the lack of observations able to constraint this
quantity (see for example Fischer et al. 1992, and van den Bergh
1999).  The present results fix this question at least down to $\sim
0.7\, M_{\sun}$.

Although the determination of the mass-to-light ratio is not one
of the scientific goals of this paper, we performed a set of
simple calculations to derive the expected value by using our
theoretical framework. For canonical models, adopting the age of
$140\, Myr$ and a Salpeter mass function in the observed range
of magnitudes, we predict $M$/L$_V$ values ranging from 0.17 to
0.26 (in solar units) depending on the assumption on the mass
distribution function in the interval 0.1$<$$M/M_{\sun}$$<$0.5.
The higher $M$/L$_V$ value corresponds to use the Salpeter mass
function down to $0.1\, M_{\sun}$, while the lower value
corresponds to a flatter Scalo function (Scalo 1986).  In the case
of overshooting, the $200\, Myr$ best fit leads to larger values
(0.20 and 0.30 respectively). As a conclusion, one finds that
cluster stars experiencing nuclear burning evolutionary phases can
hardly account for a $M$/L$_V$ as large as 0.42. If this value
will be confirmed, one cannot escape the evidence for an
additional contribution to the total mass, as given by brown
dwarfs or neutron stars (van den Bergh 1999).

\section{Overshooting or classical models?}

For the last two decades NGC~1866 has been often used to discuss
possible evidence for the occurrence of core overshooting in the
evolution of intermediate mass stars. Owing to the controversial
results appearing in the literature we will first  briefly summarize the
interpretation of previous observations.

Becker and Mathews (1983: BM83) found a scarcity of He-burning giants in
NGC~1866 with respect to their standard model predictions. However,
Brocato and Castellani (1988) warned against making premature
conclusions: the observational sample used by BM83 was severely
incomplete, reaching only to $V=18$ thus revealing only the top one
magnitude of the MS, and
possibly contaminated by binaries.
 In addition,  the BM83
stellar models did not account for either canonical semi-convection or
overshooting. New $V,B-V$ photometry down to $V=21$, including 53
He-burning giants, was presented by B89,
who used improved stellar models to show that adopting the same
distance modulus and the same metallicity as in BM83,  i.e.
$(m-M)_0=18.6$ and $Z=Z_\odot$ respectively, classical models can
indeed account for the observations. However, the same year
C89 presented independent observations of the
cluster, producing a different LF, with a significantly larger
ratio between MS and He-burning giants stars, which was taken as
an evidence against standard model and in favor of overshooting.

Both samples were limited in numbers of stars, so to solve this
problem Brocato et al. (1994: B94) took advantage of new ESO-NTT
data to explore a much larger cluster region, which contained a
total of 153 He-burning giants.  The regions covered by the
previous investigations were reanalyzed, confirming the different
results, whereas the whole sample gave a LF in good agreement with
B89. B94 concluded that the C89 LF was an artifact of statistical
fluctuations, connected to the small number of RG stars (39) in
the sample. This conclusion was further supported by the T99
reinvestigation, which with an even larger number of He-burning
giants found again a LF in very good agreement with B89,
concluding that classical models plus 30\% of binaries nicely
accounted for the observations. However, by adopting $Z = 0.008$
for the cluster metallicity, BGC02 reanalyzed the T99 dataset and
interpreted their results as evidence supporting core
overshooting, and again invoking the occurrence of a large
fraction of binary stars.

To investigate this problem,  Fig.~12, left panel compares the present LF
with theoretical expectations as given for the observed 170 RG stars by
models with or without overshooting. One finds that both models appear
at about 4 sigma from experimental data. Numerical experiments
disclose that canonical model now misses the fitting just as a
consequence of the new ``metal poor'' composition, whereas the
assumptions on the cluster distance modulus only affects the predicted
cluster ages. However, the same figure shows that the presence of 40\%
of binaries, distributed as BGC02,
moves theoretical
expectations in such a way to improve the fit with canonical models
with overshooting predictions becoming worse. As a result, the right
panel in the same figure shows that with the quoted amount of binaries
and allowing for a 2 sigma fluctuations on the number of RG stars,
canonical models do produce a reasonable fitting. Thus, we conclude that
no clear evidence against the
canonical model can be found in NGC~1866.

\section{Final remarks}

The occurrence of uncertainties and differences in the evolutionary
predictions has been often discussed in the literature, and
in particular not negligible
differences in the He-burning models as computed with
our or with the Padua code
has been already presented and exhaustively discussed in Castellani et
al. (2000).
For the case of NGC~1866 we compare in Table~3 the
evolutionary times during H and He-burning phases for classical models
with selected stellar masses covering the range of cluster off-MS
stars.

It appears that both computations give quite similar H burning
lifetimes, but differ in the He-burning phase, with Padua He
burning lifetimes larger than in our predictions. The
straightforward conclusion is that Padua computations for classical
models will predict a larger number of He-burning giants and that -- in
that case -- the introduction of overshooting as a free parameter
can help in reducing the number ratio of He-burning to MS stars.

As repeatedly discussed (see, e.g., Castellani 1999) these differences
originate in the use of different input physics; we cannot claim that our
physics is ``better'', but only that our physics is fully documented in
the recent literature.  Thus the most reasonable conclusion of this
paper is that the new {\it global} LF can be put in agreement with our
canonical model, but unfortunately NGC~1866 is in itself not
sufficient to solve the controversy of whether
or not overshooting is significant
for these stars.

\section{Acknowledgments}

This paper is based on observations made with the NASA/ESA Hubble
Space Telescope, obtained  at the Space Telescope Science
Institute, which is operated by the Association of Universities for
Research in Astronomy, Inc. under NASA contract NAS 5-26555.
Financial support for this work was provided by MIUR-Cofin 2000, under
the scientific project "Stellar Observables of Cosmological
Relevance". This project made use of computational resources granted
by the Consorzio di Ricerca del Gran Sasso according to the Progetto 6
'Calcolo Evoluto e sue Applicazioni (RSV6)' - Cluster C11/B.



\clearpage
\begin{table*}
\begin{center}
\caption{Log of the observations. Coordinates refer to center of the PC and dataset names
refer to HST archive catalog. \label{tab1}}
\begin{tabular}{ccclccc}
\hline \hline
                &   dataset  & Filter      & Exposure                &  RA                   &  Dec                  &  Position angle \\
                &   (name)   &             & time (s)                & (2000)                &  (2000)              &   (deg)         \\
\hline
PC - centered   &  u5dp510   & {\it F555W} &  $4\, \times\, 8$       & $0.5:13:43.28$        &   $-65:28:12.3$       &  126.389         \\
                &            &             &  $8\, \times\, 60$      &                       &                       &                   \\
                &            & {\it F814W} &  $4\, \times\, 5$       &                       &                       &                      \\
                &            &             &  $8\, \times\, 50$      &                       &                       &                      \\
\hline
WF3 - centered  &  u5dp020   & {\it F555W} &  $4\, \times\, 8$       & $0.5:13:33.07$        &   $-65:27:27.8$       &  125.311          \\
                &            &             &  $8\, \times\, 60$         &                       &                       &           \\
                &            &             &  $4\, \times\, 500$        &                       &                       &            \\
                &            & {\it F814W} &  $4\, \times\, 5$          &                       &                       &           \\
                &            &             &  $8\, \times\, 50$        &                       &                       &           \\
                &            &             &  $2\, \times\, 500$       &                       &                       &            \\
                &            &             &  $2\, \times\, 600$       &                       &                       &           \\
\hline \hline
\end{tabular}
\end{center}
\end{table*}

\begin{table*}[t]
\begin{center}
\caption{Mean masses of He-burning giants and masses
of MS stars at $M_V$=0 or 4.5 for the labelled assumptions on the
cluster age and for the two different assumptions about the
efficiency of overshooting. Masses in solar masses, ages in $Myr$
$Z=0.007$, $Y=0.25$.}
\vspace{0.5cm}
\begin{tabular}{cccc}
\hline \hline
\multicolumn{4}{c}{Classical models}  \\
\hline
 $AGE$      &  $M(He)$      & $M(M_V=0)$    & $M(M_V=4.5)$ \\
 (Myr)      & ($M_{\sun}$)  & ($M_{\sun}$)  & ($M_{\sun}$) \\
\hline
100    & 4.75  & 3.45  & 1.05  \\
140    & 4.15  & 3.30  & 1.05   \\
180    & 3.70  & 3.10  & 1.05  \\
\hline
\multicolumn{4}{c}{Overshooting models}  \\
\hline
160    &4.20  & 3.30  & 1.10 \\
200    &3.80  & 3.15  & 1.10 \\
250    &3.50   &3.00  & 1.10\\
\hline \hline
\end{tabular}
\label{tab2}
\end{center}
\end{table*}

\begin{table*}[t]
\begin{center}
\caption {selected evolutionary timescales for stars with
different mass as provided by the Padua group (BGC02) and the
present stellar models (labelled as Pisa). \label{tab3}}
\vspace{0.5cm}
\begin{tabular}{cccc}
\hline \hline
M              &    $t_H$     & $t_{He}$    & Models \\
($M_{\sun}$)   &    (Myr)     & (Myr)       &        \\
\hline
3.5  & 176.9 &   68.0  & Padua \\
 -   & 175.5 &   54.6  & Pisa \\
4.0  & 127.1 &   43.2  & Padua \\
  -  & 127.7 &   35.1  & Pisa  \\
5.0  & 77.5  &   21.9  & Padua \\
 -   & 77.2   &  18.8  & Pisa \\
\hline \hline
\end{tabular}
\end{center}
\end{table*}

\clearpage


\begin{figure*}[t]
\figcaption{The LMC cluster NGC~1866 as observed with
WFPC2@HST. {\it (a: Left Panel)} The PC-centered pointing
for the $50\, s$ exposures
with the filter {\it F814W} is shown. {\it (b: Right Panel)}
As in the left panel
 but the WF3-centered  pointing for the $50\, s$ exposures
with the filter {\it F814W} is shown. }
\end{figure*}

\begin{figure*}[t]
\figcaption{The present explored field of NGC~1866 is compared to
Fig.~7 by T99 where their and previous observations
(B89, C89, and B94) from ground telescopes are plotted.}
\end{figure*}

\begin{figure*}[t]
\figcaption{Upper panels ({\it a,b}) refer to the PC-centered
dataset before {\it (a: left)} and after applying
all the corrective procedures {\it (b: right)}.
Lower panels ({\it c,d}) the same but for the WF3-centered CMDs.}
\end{figure*}

\begin{figure*}[t]
       \caption{
Uncertainties of PC-centered photometry as derived by HSTphot
for the {\it F555W} {\it (a)} and
the {\it F814W} bands {\it (b)}.}
\end{figure*}

\begin{figure*}
\figcaption{Completeness factor as obtained with the
PC-centered {\it (a: left panel)}
and WF3-centered {\it (b: right panel)} datasets
for the labelled annuli. The curves obtained for
the two photometric bands are shown.}
\end{figure*}

\begin{figure*}[t]
\caption{The differential luminosity function normalized to the
proper number of  He-burning giants (RG) is shown at different
distance from the cluster center. The dashed lines represent the
raw data after all the photometric corrections (see text), the
dotted lines show the LFs after the completeness corrections
and the solid lines are the final LFs as obtained by subtracting
the expected contamination of field stars.}
\end{figure*}

\begin{figure*}[t]
\figcaption{The differential LF of the PC (thin solid line) and
WF3 (dotted line)
datasets are compared after corrections due to completeness and
field contaminations.
The {\it global} LF (bold solid line) as derived by combining
the observed regions of
the two datasets is also shown. All the LFs are normalized to
the proper number of RG stars
(i.e. 128, 153, and 170, respectively). The uncertainty of the
{\it global} LF
 due to the number of RG stars is plotted in the right bottom.}
\end{figure*}

\begin{figure*}[t]
\figcaption{The {\it global} LF is compared to the LFs derived
in previous papers.
The number of RG stars is also reported.}
\end{figure*}

\begin{figure*}[t]
\caption{The CMD of the PC-centered dataset is reported (blu dots)
and compared
with the synthetic CMDs (red dots) computed by adopting {\it classical}
{\it (left panels)} and {\it mild-overshooting} {\it (right panels)}
stellar models for the labelled ages.
\label{fig10}}
\end{figure*}

\begin{figure*}[t]
\figcaption{The log of
the theoretical LFs derived for the labelled ages is reported.
The upper panel
refers to the {\it classical} stellar models, the lower panel to the
{\it mild-overshooting} stellar models.}
\end{figure*}

\begin{figure*}[t]
\figcaption{The {\it global} LF (red line) is compared to theoretical LFs as
derived by assuming
the labelled values of the Salpeter exponent ($\alpha = 1 + x$)
in the quoted range of masses. The solid black line refers to the value
$\alpha = 2.35$.}
\end{figure*}

\begin{figure*}[t]
\figcaption{{\it (a: Left Panel)}
The {\it global} LF (bold solid line) is compared to theoretical LFs. The
dashed line refers to the LF derived by {\it mild-overshooting} models.
The thin solid line is the LF predicted by  {\it classical} models
and the dotted line is the LF expected by  {\it classical}
models with the contribution of 40\% of binaries (see text). {\it
(b: Right Panel)} As in the left panel  but the dotted line is obtained
allowing a 2 sigma fluctuations of the He-burning giant stars from
the LF expected by  {\it classical} models with the contribution of
40\% of binaries (see text). }
\end{figure*}

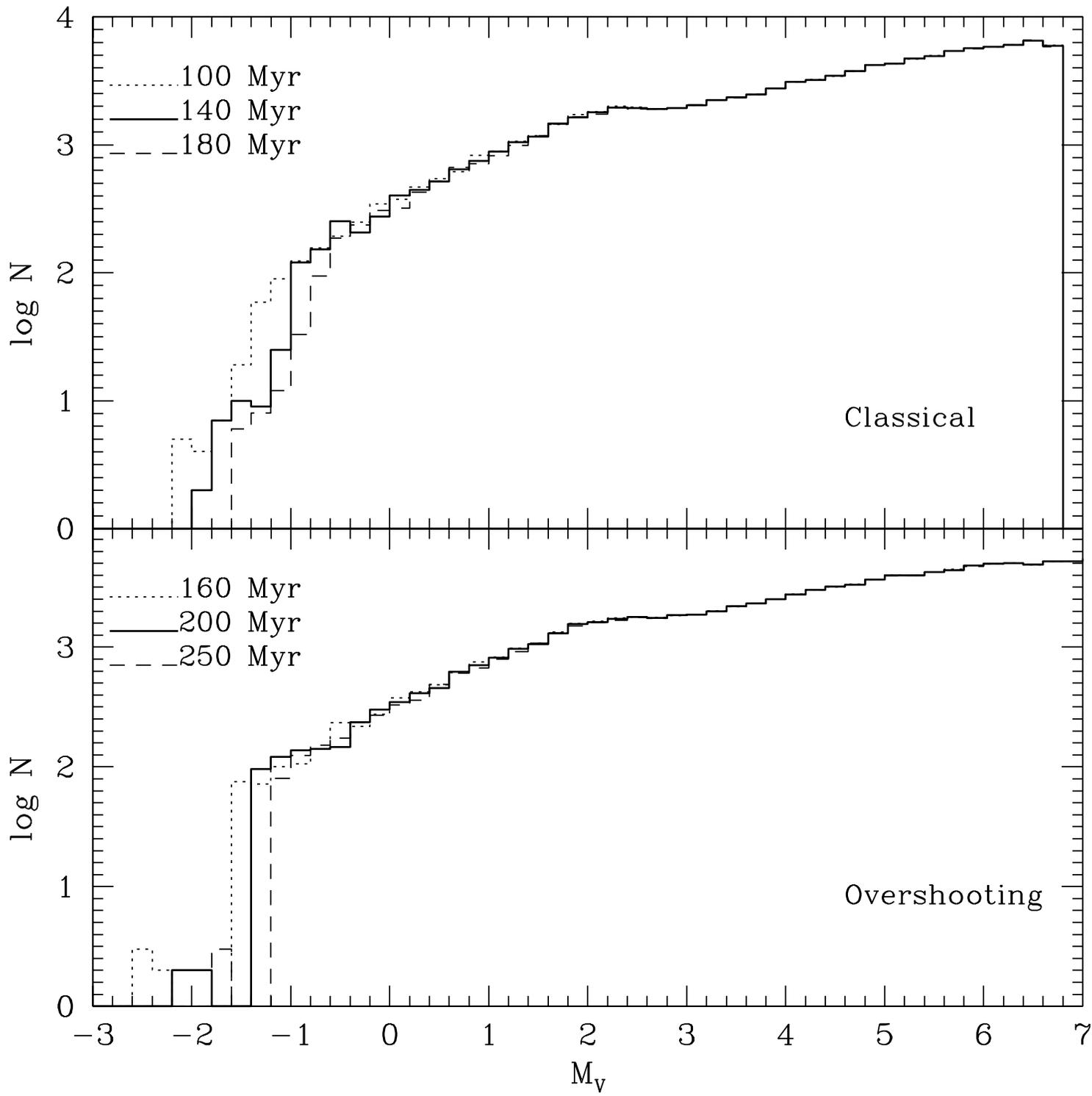
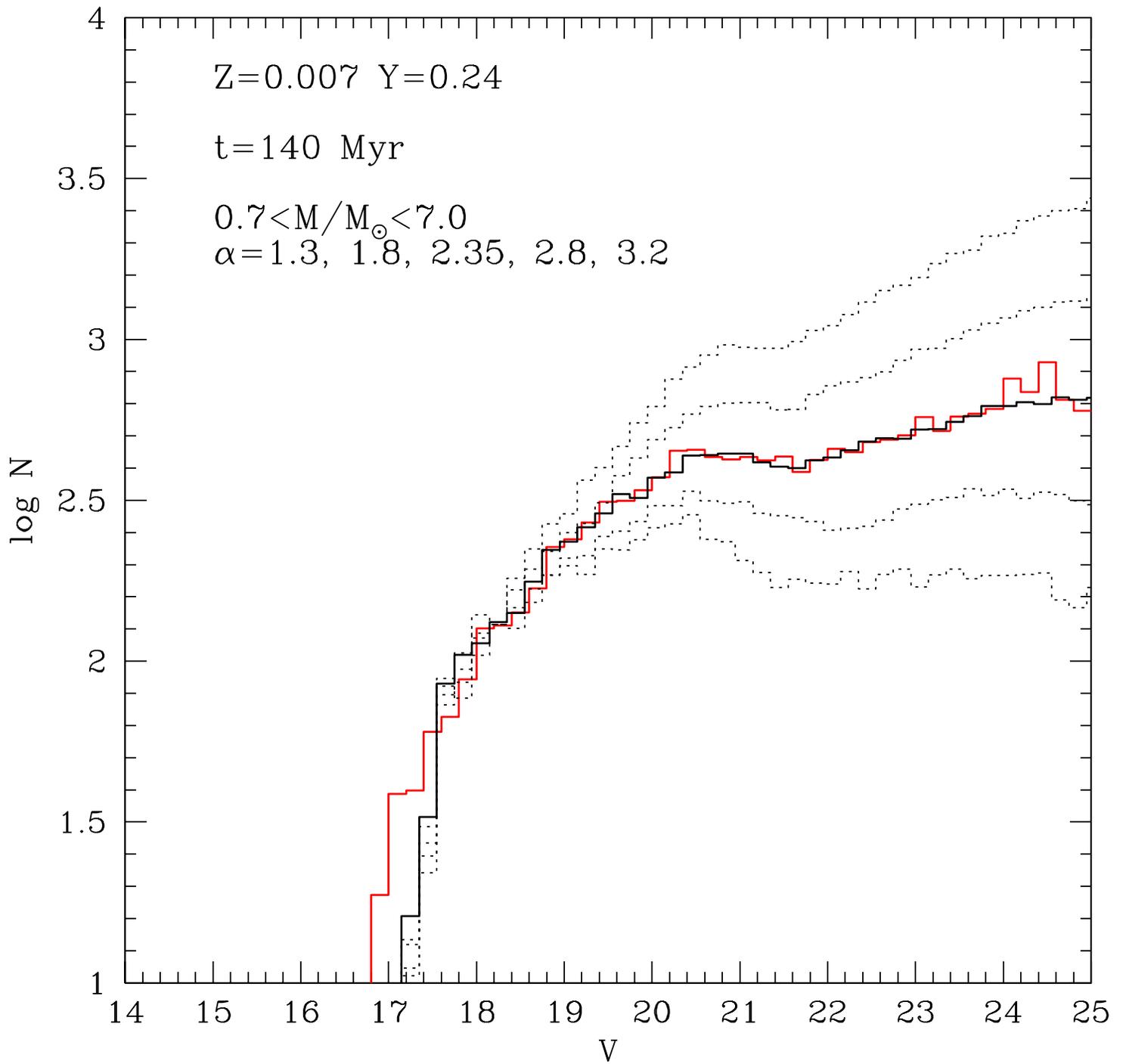
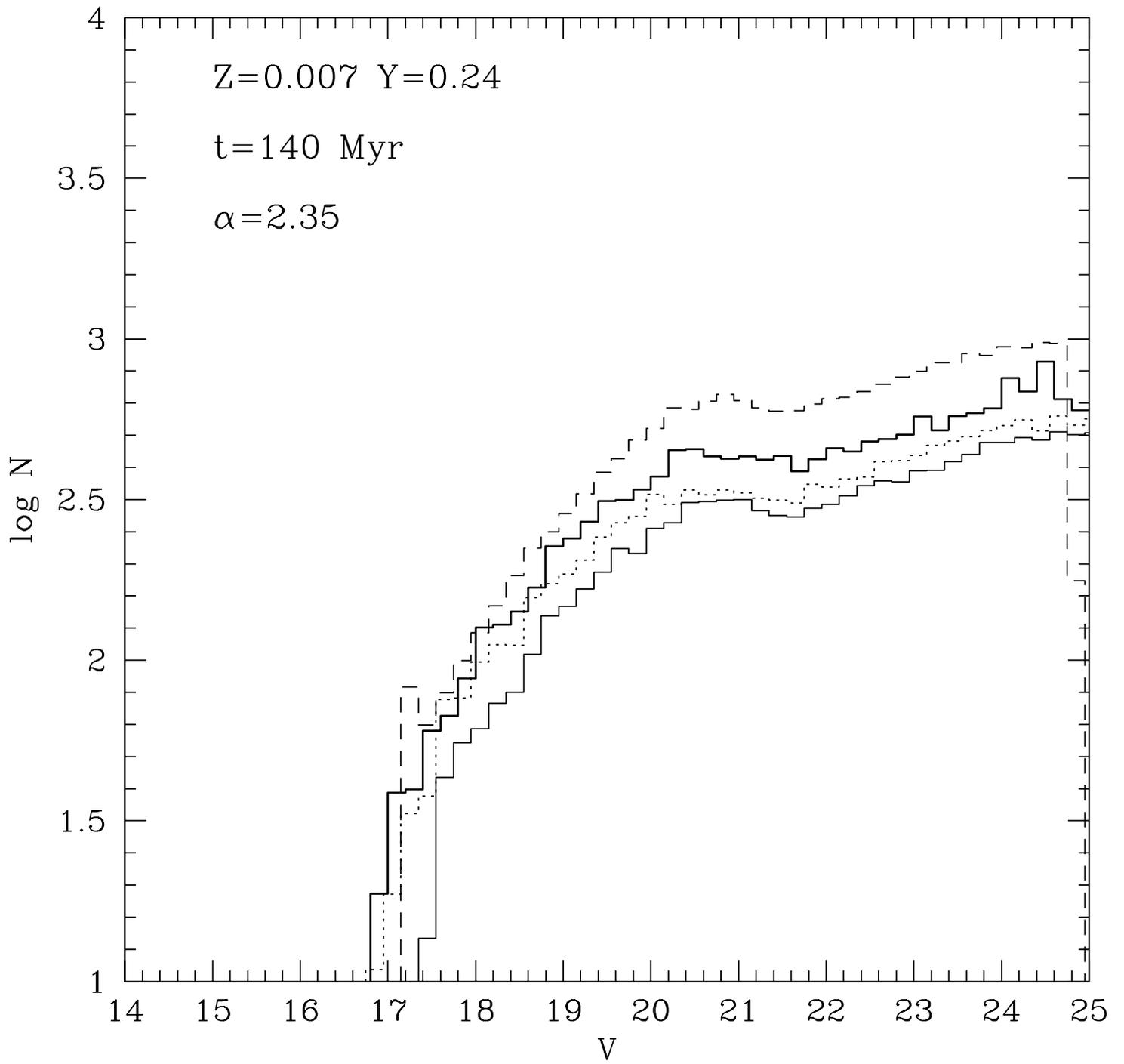
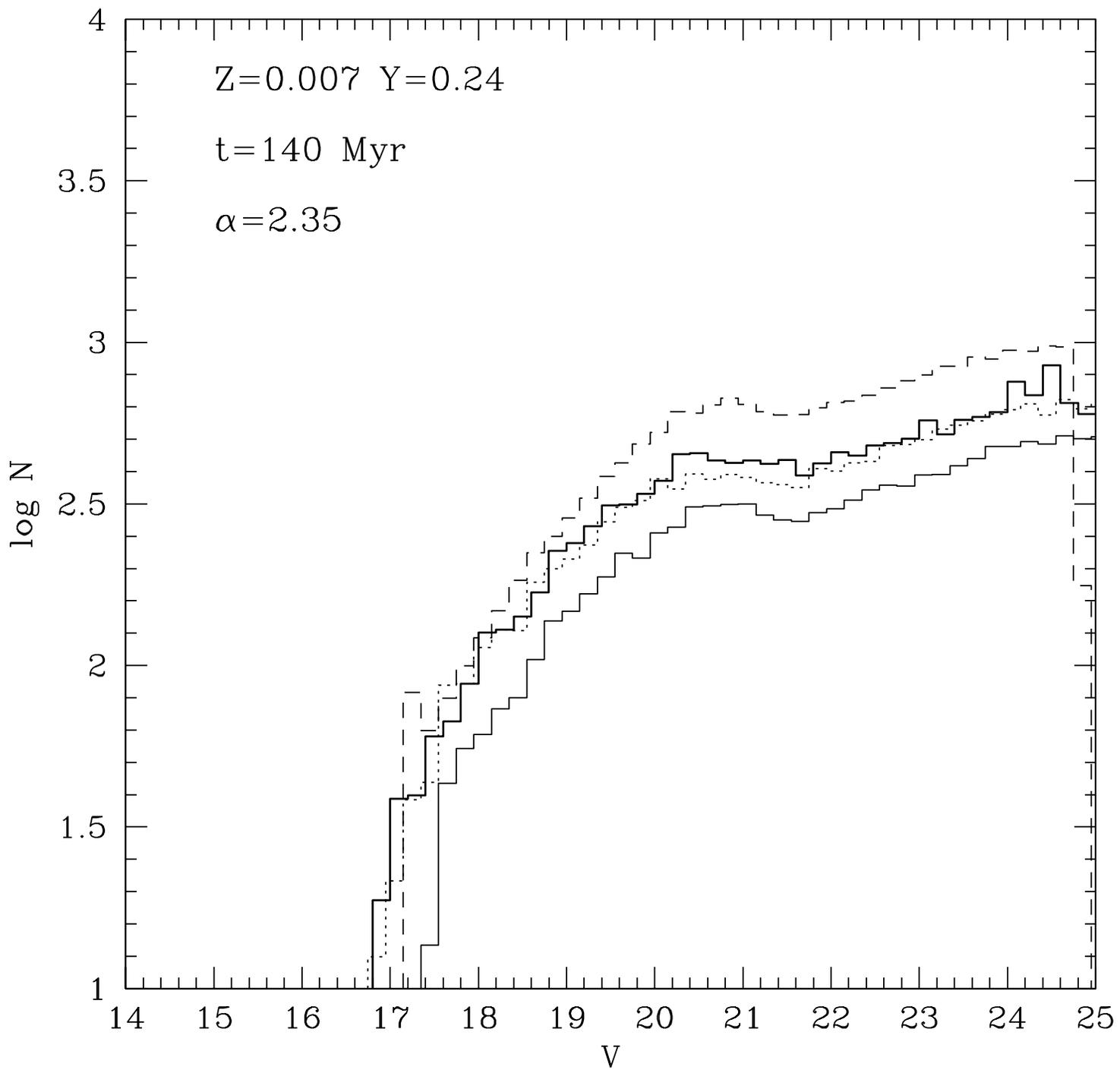
\end{document}